# Optical Leaky-Wave Antenna Integrated in Ring Resonator

Caner Guclu, Mehdi Veysi, Ozdal Boyraz, and Filippo Capolino
Department of Electrical Engineering and Computer Science
University of California, Irvine
Irvine, California, United States
f.capolino@uci.edu

*Abstract*—A leaky-wave antenna at optical frequencies is designed and integrated with a ring resonator at 1550 nm wavelength. The leaky wave is generated by using periodic perturbations in the integrated dielectric waveguide that excite the -1 spatial harmonic. The antenna consists of a dielectric waveguides with semiconductor corrugations, and it is compatible with CMOS fabrication technology. We show that integrating the leaky wave antenna in an optical ring resonator that is fed by directional couplers, we can improve the electronic control of the radiation through carrier injection into the semiconductor corrugations.

## I. Introduction

Optical leaky-wave antennas (OLWAs) have been presented in [1]-[3] where a limited radiation control mechanism of optical pumping was demonstrated in [2] and [3]. In order to enhance the control of the radiation level we have proposed to integrate the OLWA in a Fabry-Pérot resonator (FPR) [4] and [5]. For details on the topic, readers are referred to the references in [1]-[5]. Other examples of OLWAs include an antenna comprising a groove in aluminum superstrate on top of a silver substrate demonstrated in [6], and linear chains of plasmonic nanoparticles that support leaky modes as shown in [7] and also in [8] with an extensive mode characterization.

The OLWA integrated in FPR was first conceptually designed and its performance in a two dimensional (2D) design was demonstrated in [5]. On the other hand, as stated therein, the difficulties of implementing and optimizing mirrors in silicon-on-insulator (SOI) waveguides motivated our study to integrate the OLWA in a ring resonator. Moreover, here we will further analyze the impact of three-dimensional design in contrast to the 2D one shown in [5]. The ultimate aim is to achieve radiation level control through modulating the excess carriers in Si domain of the antenna.

## II. Analytical Model and Results

In this section the wave dynamics in a ring resonator with a radiating section is analytically investigated and an example of far-field modulation is demonstrated. In Fig. 1, we provide the illustration of the resonator with a racetrack geometry fed through a directional dielectric waveguide coupler. The radiating section is composed of a dielectric waveguide with periodic perturbations, with period $d$, which lead to the excitation of the radiating -1 Floquet spatial harmonic. When the period is chosen very close to the guided wavelength and very small perturbations are used, the wavenumber of -1 harmonic $k_{\mathrm{LW}} = k - 2\pi/d$ can be adjusted as a leaky-wave with wavenumber $k_{\mathrm{LW}} = \beta_{\mathrm{LW}} + i\alpha_{\mathrm{LW}}$ which in turn leads to a directive radiated beam close to the broadside direction. The details of OLWA design are reported in [1].

The waveguide is fed at the input with $y$-polarized electric field. The electromagnetic wave coupled to the resonator will circulate in counter-clockwise direction. At frequencies where $m \times 2\pi$ phase accumulation occurs (where $m$ is an integer) upon a roundtrip, the resonance occurs leading to a large field in the resonator. When material and radiation losses are low the ring supports a very high quality resonance. Therefore leaky-wave radiation can occur over a narrow frequency region, this implies that the resonator can be pushed out of resonance with a small variation of the propagating mode's wavenumber. The radiated beam for the racetrack resonator in Fig. 1 (which hosts a leaky wave propagating along the $x$-direction) is directive in the $xz$ plane whereas in the $yz$ plane the pattern is similar to the far-field of a $y$-directed dipole. The radiated beam exhibits a fan-shaped pattern as illustrated in Fig. 1. More complicated OLWA in ring resonator topologies with several antenna segments and different shapes of the resonator can be designed based on the required far-field pattern and polarization. For demonstration purposes we report here only the simple racetrack topology in Fig. 1.

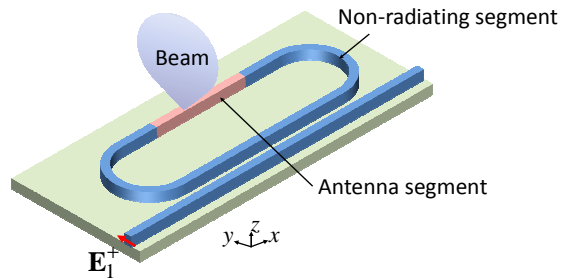

Fig. 1. The illustration of OLWA integrated with ring resonators.

Wave dynamics and resonance conditions in the ring resonator are shown in Fig. 2. The coupling coefficient and the transmission coefficient of the directional coupler for the electric field are $\kappa = 0.3$ and $t = 0.95$, respectively. Here, we assume that the output port is matched to the dielectric waveguide, thus there is no reflection i.e, $E_2^- = 0$.

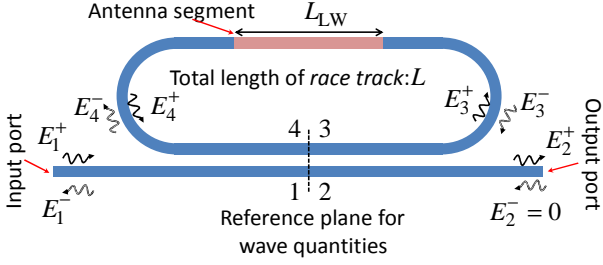

Fig. 2. Wave dynamics of OLWA in a racetrack resonator.

In general, the ring resonator hosts waves that propagate both in the clockwise and counter-clockwise directions, therefore the radiation in the antenna segments is due to two counter propagating waves. This is a very similar situation as discussed in [5] for 2D OLWA in a FPR. For simplicity, here we assume that the antenna segment (the radiating part) is well matched with the non-radiating segments (which is assumed lossless), therefore only counter-clockwise propagating waves will be excited in the ring. The resonance condition occurs when $\beta_{LW} L_{LW} + k(L - L_{LW}) + \angle t = 2\pi m$ where $m$ is an integer. Then the E-field in resonator is given by

$$E_3^+ = E_{LW}^+ = E_1^+ \frac{i\kappa e^{ik_{LW} L_{LW}/2} e^{ik(L-L_{LW})/2}}{1 - t e^{ik_{LW} L_{LW}} e^{ik 2(L-L_{LW})}}. \quad (1)$$

At resonance the denominator becomes very small ( $\left|1 - |t| e^{-\alpha_{LW} L_{LW}}\right|$ ) leading to enhanced fields in the ring resonator. Next we demonstrate analytically how the enhanced fields can be controlled. Assuming absence of excess carriers in a Si waveguide, the wavenumber of the leaky wave is $k_{LW} = (-0.0406 + i0.0014)k_0$ [ $k_0$ is the wavenumber in vacuum], the antenna segment's length is $L_{LW} = 5.98\,\mu\text{m}$, the phase accumulated in the non-radiating segment is $k(L - L_{LW}) = (40.15)2\pi$ based on full-wave simulations. For this case, the attenuation due to the radiation in the antenna segment, the only loss mechanism, becomes $e^{-\alpha_{LW} L_{LW}} = 0.97$ at 1550 nm wavelength. We denote the absence of excess carriers as "State 1". Injection of excess carriers with density $N_e = 10^{19}\,\text{cm}^{-3}$ (denoted as "State 2") in the antenna segment leads to a slight change in real part of the leak wave's wavenumber as $k_{LW} = (-0.0205 + i0.0072)k_0$ whereas increased material losses in Si lead to an attenuation of $e^{-\alpha_{LW} L_{LW}} = 0.84$, as discussed in [5]. State 2 pushes the resonator out of resonance at 1550 nm as shown in Fig. 3(a). Moreover the change in the leaky wave amplitude $E_{LW}^+$ directly translates to a radiation level change as shown in Fig. 3(b) where the far-field radiation pattern is computed using the analytical formula provided in [5]. By this analytical approach we provide a way to modulate the far-field radiation level of an OLWA in a ring resonator by 15 dB.

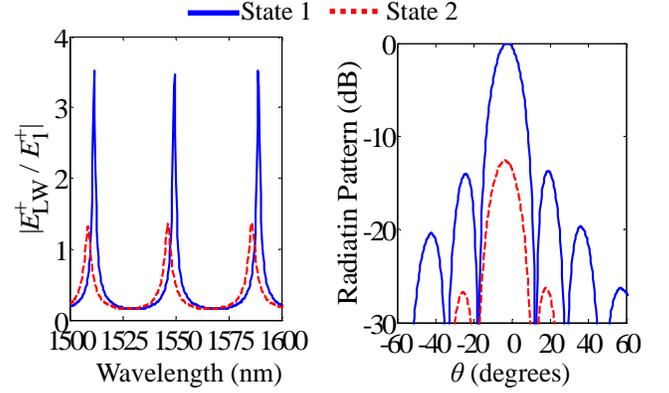

Fig. 3. (a) The field magnitude of the leaky wave in the antenna section versus wavelength. (b) Far-field radiation pattern in the xz plane versus $\theta$, the angle with the z-axis, at 1550 nm wavelength, normalized by the maximum of "State 1" condition.

The development of sharply resonant tunable OLWAs integrated in resonators opens up the possibilities for fast optical switches and frequency-sensitive sensors at optical frequencies. The design robustness to fabrication tolerances and experimental verifications of the analytical approach are yet to be explored.

ACKNOWLEDGEMENT

This research is partially funded by the National Science Foundation under the NSF award #ECCS-1028727.